\documentclass[%twocolumn,
secnumarabic, amssymb, amsmath, nofootinbib,tightenlines, showpacs, nobibnotes, aps, prl]{revtex4}
\usepackage{amsfonts}

\usepackage{docs}%
\usepackage{bm}%
\usepackage[colorlinks=true,linkcolor=blue]{hyperref}%
\usepackage{graphicx}% Include figure files
\usepackage{dcolumn}% Align table columns on decimal point

\begin{document}

\title{BCS-BEC Crossover in Mix-dimensional Fermi Gases}

\author{Xiaosen Yang$^{1}$}
\author{Beibing Huang$^{2}$}
\author{Shaolong Wan$^{1}$}
\altaffiliation{Corresponding author} \email{slwan@ustc.edu.cn}
\affiliation{$^{1}$Institute for Theoretical Physics and Department of Modern Physics \\
University of Science and Technology of China, Hefei, 230026, {\bf
P. R. China} \\
$^{2}$Department of Experiment Teaching, Yancheng
Institute of Technology, Yancheng, 224051, {\bf P. R. China}}

\date{\today}

\begin{abstract}
We investigate a mix-dimensional Fermi-Fermi mixture in which one
species is confined in two-dimensional(2D) space while the other
is free in three-dimensional space(3D). We determine the
superfluid transition temperature $T_{c}$ for the entire BCS-BEC
crossover including the important effects of noncondensed pairs.
We find that the transition temperature reduces while the
imbalance of mass is increased or lattice constant ($dz$) is
reduced. In population imbalance case, the stability of superfluid
is sharply destroyed by increasing the polarization.
\end{abstract}


\maketitle

\section{Introduction}

Ultracold Fermi gases have attracted considerable attention
\cite{Schunck, Zwierlein, Simons, Recati, Beibing Huang, Giorgini,
Bloch} with their various tunability like the interaction strength
via the Feshbach resonance and the dimensions of the system by
means of optical lattices. This provides a wonderful opportunity
to simulate and study the superfluidity of the fermions for entire
BCS-BEC crossover. Lately, the population imbalance between the
two species can also be varied. Since the imbalance causing the
Fermi surface mismatch destabilizes pairing between the two
species and enriches the ground state of the system \cite{Simons,
Recati, Qijin chen1, Qijin chen2, Bedaque, Forbes, Wei Zhang,
Chevy, Gubbels}. The imbalance system possess exotic superfluid
phases such as Sarma phase \cite{Gubbels} mixture of the normal
and superfluid phase, Larkin-Ovchinnikov-Fulde-Ferrel (LOFF) state
\cite{Wei Zhang, Yan He} in which the condensed pairs have nonzero
net momentum.

Recently, the mix-dimensional system where the different species
live in different spatial dimensions \cite{Nishida, Nishida1,
iskin, Young-S} has been realized with Bose-Bose mixture
\cite{G.Lamporesi} by exploiting a species-selective 1D optical
lattice to confine only one atomic species($^{41}K$) in 2D while
the other ($^{87}Rb$) is free in the 3D space. In the
mix-dimensional system, the Efimov effect \cite{Kraemer, Lee,
Yusuke Nishida} takes place only in some range of mass ratio for
Fermi-Fermi mixtures \cite{Nishida} while for any mass ratio for
Bose-Bose and Bose-Fermi mixtures at resonance. The
mix-dimensional systems open a new subfield in cold atom for
investigating the heteronuclear molecules whose two species
constituents live in different dimensions.

In this paper we consider the two-species Fermi-Fermi mixtures in
the mix-dimensional(2D-3D) system and obtain the behavior of the
superfluid transition temperatures $T_{c}$ for the entire BCS-BEC
crossover with different optical lattice parameters, the
intra-species mass ratio and polarization. In addition we give the
behavior of $T_{c}$ in resonance limit. We consider only a
uniformly superfluid of the mixtures excluding from consideration
the phase separated state. As for the two-species living in
different spatial dimensions, the Fermi surfaces of the
two-species are mismatched which also can be tuned by the
parameters of the optical lattice, mass ratio and the
polarizations. We analyze the superfluid of the mix-dimensional
system with mismatched Fermi surfaces including the the effects of
noncondensed pairs which is called "pseudogap effects" \cite{Qijin
chen, C. C. Chien} based on the BCS-Leggett ground state. For the
nonzero temperature, the excitation gap($\Delta$) contains two
contributions from pseudogap for the noncondensed
pairs($\Delta_{pg}$) and the order parameter($\Delta_{sc}$). The
pseudogap contribution vanishes at zero temperature while the
order parameter vanish at the critical transition temperature
$T_{c}$. The main conclusions are as follows: (a) increasing of
the optical lattice parameter $dz$ (half wave length of the
optical lattice) can improve the transition temperature; (b) the
masses and the popularizations imbalances of the two-species both
suppress the transition temperature $T_{c}$; (c) in population
imbalance case, the superfluid phase is in a narrow region($- 0.2
< p < 0.2$) and unstable for high polarization.

This paper is organized as follows. In Sec. II, the formalism of
the mix-dimensional fermi gases are introduced. In Sec. III, the
numerical results and discussions are given. The conclusion are
given in Sec. IV.

\section{Formalism of the Mix-dimensional Fermi Gases}

The Hamiltonian for the mix-dimensional Fermi-Fermi mixtures
system can be written as($\hbar = k_{B} = 1$)
\begin{eqnarray}
H = \sum_{\textbf{k}, \sigma} \xi_{\textbf{k}, \sigma}
a_{\textbf{k}, \sigma}^{\dag} a_{\textbf{k}, \sigma} + g
\sum_{\textbf{k}, \textbf{k'}, \textbf{q}} a_{\textbf{q/2} +
\textbf{k}, \uparrow}^{\dag} a_{\textbf{q/2} - \textbf{k},
\downarrow}^{\dag} a_{\textbf{q/2} - \textbf{k'}, \downarrow}
a_{\textbf{q/2} + \textbf{k'}, \uparrow}, \label{2.1}
\end{eqnarray}
where, the pseudo-spin $\sigma = {\uparrow, \downarrow}$ labels
the two types of Fermi atoms. We assume that the $\uparrow$ Fermi
atoms is confined in a species-selective 1D optical lattice while
the other is free in the 3D. The free dispersions of the two
species Fermi atoms are $\xi_{\textbf{k}, \uparrow} =
\frac{k_{x}^{2} + k_{y}^{2}}{2 m_{\uparrow}} + 2 t [1 - cos(k_{z}
d_{z})] - \mu_{\uparrow}$ and $\xi_{\textbf{k}, \downarrow} =
\frac{k_{x}^{2} + k_{y}^{2} + k_{z}^2}{2 m_{\downarrow}} -
\mu_{\downarrow}$. Here $\mu_{\sigma}$ is the two species Fermi
atoms chemical potential, $t$ is the amplitude of the $\uparrow$
Fermi atoms tunnelling to the nearest-neighbor sites, and $d_{z}$
is the lattice parameter which is half wave length of the 1D
optical lattice.

We determine the transition temperature $T_{c}$ including the pair
fluctuation (pseudogap effect). Truncating the equation of motion
for Green's functions scheme we can get the pair propagator
\begin{eqnarray}
t(Q) = \frac{g}{1 + g \chi(Q)}, \label{2.2}
\end{eqnarray}
and the self-energy
\begin{eqnarray}
\Sigma_{\sigma}(K) = \sum_{Q} t(Q) G_{0, \sigma}(Q - K),
\label{2.3}
\end{eqnarray}
where the pair susceptibility is given by
\begin{eqnarray}
\chi(Q) = \frac{1}{2}[\chi_{\uparrow \downarrow}(Q) +
\chi_{\downarrow \uparrow}(Q)] = \frac{1}{2} \sum_{K}[G_{0,
\uparrow}(Q - K) G_{\downarrow}(K) + G_{0, \downarrow}(Q - K)
G_{\uparrow}(K)], \label{2.4}
\end{eqnarray}
with the bare green's function $G_{0, \sigma}^{-1} (K) = \imath
\omega_{n} - \xi_{\textbf{k}, \sigma}$, and the $G_{\sigma} (K)$
is dressed green's function. We take the notation $K = (\imath
\omega_{n}, \textbf{k})$, $Q = (\imath \Omega_{n}, \textbf{q})$,
$\Sigma_{K} = T \Sigma_{n} \Sigma_{\textbf{k}}$, etc., where
$\omega_{n}(\Omega_{n})$ is the odd(even) Matsubara frequency.

Below the critical temperature $T_{c}$, the \textbf{T} matrix and
self energy contains both the condensed($sc$) and the noncondensed
or "pseudogap"-associated($pg$) contributions:
\begin{eqnarray}
t (Q) = t_{sc}(Q) + t_{pg}(Q), \label{2.5}
\end{eqnarray}
\begin{eqnarray}
t_{sc} (Q) = - \frac{\Delta_{sc}}{T} \delta(Q), \label{2.6}
\end{eqnarray}
\begin{eqnarray}
t_{pg} (Q) = \frac{g}{1 + g \chi(Q)},~~~~~~~~ Q \neq 0.
\label{2.7}
\end{eqnarray}
So the total fermion self-energy is given by
\begin{eqnarray}
\Sigma_{\sigma} (K) = \Sigma_{\sigma, sc}(K) + \Sigma_{\sigma,
pg}(K) = - \Delta^{2} G_{0, \overline{\sigma}}(- K), \label{2.8}
\end{eqnarray}
where the excitation gap contain two contributions of the order
parameter ($sc$) and the  pseudogap($pg$) $\Delta^{2} =
\Delta_{sc}^{2} + \Delta_{pg}^{2}$, with the pseudogap
$\Delta_{pg}^{2} = - \sum_{Q \neq 0} t_{pg}(Q)$.

The dressed green's function $G_{\sigma}$ can be derived by Dyson
equation
\begin{eqnarray}
G_{\uparrow} (K) = \frac{u_{\textbf{k}}^{2}}{\imath \omega_{n} -
\xi_{\textbf{k}}^{\alpha}} + \frac{v_{\textbf{k}}^{2}} {\imath
\omega_{n} + \xi_{\textbf{k}}^{\beta}}, \label{2.9}
\end{eqnarray}
\begin{eqnarray}
G_{\downarrow}(K) = \frac{u_{\textbf{k}}^{2}}{\imath \omega_{n} -
\xi_{\textbf{k}}^{\beta}} + \frac{v_{\textbf{k}}^{2}} {\imath
\omega_{n} + \xi_{\textbf{k}}^{\alpha}}, \label{2.10}
\end{eqnarray}
where, the fermion excitation $\xi_{\textbf{k}}^{\alpha, \beta} =
E_{\textbf{k}} \pm\xi_{\textbf{k}}^{-}$, with $E_{\textbf{k}} =
\sqrt{\xi_{\textbf{k}}^{+ 2} + \Delta^{2}}$,
$\xi_{\textbf{k}}^{\pm} = (\xi_{\textbf{k}, \uparrow} \pm
\xi_{\textbf{k}, \downarrow})/2$, $u_{\textbf{k}}^{2}  = (1 +
\xi_{\textbf{k}}^{+}/E_{\textbf{k}})/2,~ v_{\textbf{k}}^{2} = (1 -
\xi_{\textbf{k}}^{+}/E_{\textbf{k}})/2$.

In superfluid state, the "gap equation" is given by the Thouless
criterion, $t^{-1}(Q = 0) = 0$, which is equivalent to the BEC
condition of pairs, $\mu_{pair} = 0$. The coupling constant $g$ is
written in terms of the s-wave scattering length $a$, via the
relationship $M/(4 \pi a) = 1/g +
\Sigma_{\textbf{k}}(2\epsilon_{\textbf{k}}^{+})$, where $M = 2
m_{\uparrow} m_{\downarrow}/(m_{\uparrow} + m_{\downarrow})$ and
$\epsilon^{\pm} = (\epsilon_{\uparrow} \pm
\epsilon_{\uparrow})/2$. The gap equation reduce to
\begin{eqnarray}
- \frac{M}{2 \pi a} = \sum_{\textbf{k}} [\frac{1 -
f(\xi_{\textbf{k}}^{\alpha}) - f(\xi_{\textbf{k}}^{\beta})} {2
E_{\textbf{k}}} - \frac{1}{\epsilon_{\textbf{k}}^{+}}],
\label{2.11}
\end{eqnarray}
where the $f(x)$ is the Fermi distribution function. The number of
$\sigma$ fermion: $n_{\sigma} = \Sigma_{K} G_{\sigma}(K)$. The
equations of the total number $n = n_{\uparrow} + n_{\downarrow}$
and the number difference  $\emph{p} n = n_{\uparrow} -
n_{\downarrow}$ of fermions are
\begin{eqnarray}
n = \sum_{\textbf{k}}[2 v_{\textbf{k}}^{2} +
\frac{\xi_{\textbf{k}}^{+}}{E_{\textbf{k}}}
(f(\xi_{\textbf{k}}^{\alpha}) + f(\xi_{\textbf{k}}^{\beta}))],
\label{2.12}
\end{eqnarray}
\begin{eqnarray}
\emph{p} n = \sum_{\textbf{k}}[f(\xi_{\textbf{k}}^{\alpha}) -
f(\xi_{\textbf{k}}^{\beta})], \label{2.13}
\end{eqnarray}
here, $\emph{p}$ is polarization of the two-species Fermi atoms.
Since the contribution to pseudogap is dominated by the small $Q$
divergent region for the BEC condition. The $T$ matrix can be
expanded in low energy and long wavelength limit. The equation of
pseudogap can be written as
\begin{eqnarray}
\Delta_{pg}^{2} = - \sum_{Q}t_{pg}(Q) = Z^{-1} \sum_{\textbf{q}}
b(\Omega_{\textbf{q}}), \label{2.14}
\end{eqnarray}
where, $\Omega_{\textbf{q}} = \frac{\textbf{q}_{\bot}^{2}}{2
M_{\bot}^{*}} + \frac{\textbf{q}_{z}^{2}}{2 M_{z}^{*}}$ is pair
dispersion with $M_{\perp}^{*}$ and $M_{z}^{*}$ representing the
anisotropic effective pair mass computed from the low energy and
long wavelength expansion of the pair susceptibility
$\chi(Q)$[given in Appendix] and $b(\Omega_{\textbf{q}})$ is the
boson distribution.

The stability of the superfluid phase requires the positive
definiteness of the number susceptibility matrix, which is
equivalent to the positive second-order partial derivation of the
thermodynamical potential with respect to the excitation gap
$(\partial^{2} \Omega/\partial \Delta^{2})_{\mu_{\uparrow},
\mu_{\downarrow}} > 0$
\begin{eqnarray}
(\frac{\partial^{2} \Omega}{\partial \Delta^{2}})_{\mu_{\uparrow},
\mu_{\downarrow}} = 2 \sum_{\textbf{k}}
\frac{\Delta^{2}}{E_{\textbf{k}}^{2}}[\frac{1 -
f(\xi_{\textbf{k}}^{\alpha}) - f(\xi_{\textbf{k}}^{\beta})} {2
E_{\textbf{k}}} + \frac{f'(\xi_{\textbf{k}}^{\alpha}) +
f'(\xi_{\textbf{k}}^{\beta})}{2}] > 0, \label{2.15}
\end{eqnarray}
In addition, we also consider the positivity of the anisotropic
effective pair mass $M_{z}^{*}$ and $M_{\perp}^{*}$ .
Eqs.(\ref{2.11})-(\ref{2.14}) are closed for our mix-dimensional
system. For $T = 0$, the pseudogap contribution vanish so the
excitation gap $(\Delta = \Delta_{sc})$. For $T > T_{c}$, the
order parameter $(\Delta_{sc})$ vanish such that the excitation
gap contains the pseudogap only $(\Delta = \Delta_{pg})$. The
transition temperature $T_{c}$ can be determined by
self-consistently solving this closed equations throughout the
entire BCS-BEC crossover with $\Delta_{sc}^{2} = 0$.

\section{Numerical Results and Discussions}

First, the transition temperature $T_{c}$, are obtained from the
self-consistent equations, as a function of scattering length,
$1/k_{F} a$, for different optical lattice parameter, $k_{F} dz$,
with the tunnelling amplitude $t = E_{F}$ in population symmetry
case ($p=0$), shown in Fig.[1]. In the BCS limit, $1/k_{F} a
\rightarrow - \infty$, the transition temperature $T_{c}$ as well
as the anisotropic effective mass $M_{z}$ and $M_{\perp}$ of the
cooper pairs is reduce to zero. As the interaction increase to the
unitarity, $T_{c}$ rises rapidly. In the BEC limit, $1/k_{F} a
\rightarrow \infty$, $T_{c}$ varies very slowly and the
anisotropic effective mass of the pairs $M^{*}_{\perp} =  2m$
while $M^{*}_{z}$ depends on the value of $k_{F} dz$. Increasing
the parameter $dz$, lattice constant of the 1D optical lattice,
improves the transition temperature for the entire BCS-BEC
crossover. This phenomenon can be illustrated as follows. When the
tunnelling amplitude $t=E_{F}$, the Fermi surface of the
$\downarrow$ is a spherical shell while the $\uparrow$ is a
prolate spheroid. Increasing of the parameter $k_{F}dz$ can
decrease the mismatch of the two Fermi surface such that
increasing $k_{F} dz$ is propitious to form the cooper pairs. If
the parameter $dz$ is much larger than the inverse of Fermi
momentum $k_{F}^{-1}$, the localization of the $\uparrow$ fermion
in $\hat{\textbf{z}}$ direction is very weak and the
Mix-dimensional effect is reduced. Hence the parameter $dz$ must
not be much larger than the inverse of Fermi momentum
$k_{F}^{-1}$.

Second, the transition temperature $T_{c}$, are obtained from the
self-consistent equations, as a function of $1/k_{F} a$ for
different mass ratio $\eta$, shown in fig.[2]. We consider the
population symmetry, the tunnelling amplitude $t=E_{F}$ and the
parameter $k_{F} dz = 1$. In terms of the mixed dimensions system
of Fermi-Fermi mixture(2D-3D), the system is stable against the
Efimov effect\cite{Nishida, Lee, Yusuke Nishida} for even-parity
case.

Then, the transition temperature $T_{c}$ attains the maximum at
the mass balance case and reduces for the mass ratio deviating
from the symmetry case as shown in the Fig.[3] at resonance limit.
The imbalance of the mass disturbs the inter-species pairing.

Finally, for the population imbalance case, we find that the
superfluid phase is restricted in the region of low polarization
$- 0.2 < p < 0.2$ at resonance, shown in Fig.[4]. The transition
temperature $T_{c}$ reduces while the polarization is increased.
The superfluid phase in the BEC limit is not stable for the
condition of stability is violated as $(\partial^{2}
\Omega/\partial \Delta^{2})_{\mu_{\uparrow}, \mu_{\downarrow}} <
0$. At high polarization, the superfluid is unstable in the entire
BCS-BEC region which is very different from the pure three
dimensions systems \cite{Hao guo}. The superfluid of the
Fermi-Fermi mixture system in mixed dimensions is very unstable
with population imbalance. In BCS side the anisotropic effective
mass smoothly reduce to zero. In the unstable regions, other
possible phase like the LOFF state maybe occur which will be
consider in the future works.

\begin{figure}
\includegraphics[width=8.5cm, height=6.0cm]{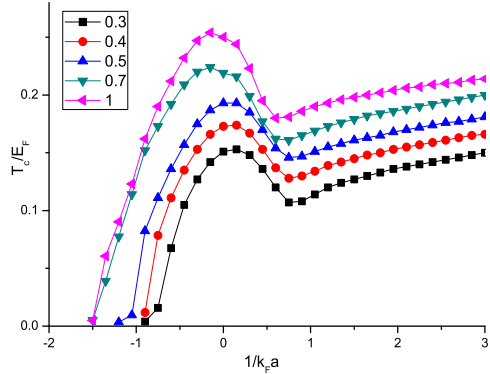}
\caption{(color online). $T_{c}$ as a function of $1/k_{F} a$ for
different lattice parameter $k_{F} dz$ with the polarization $p =
0$; $t = E_{F} = \hbar^{2} k_{F}^{2}/2 m$ and
$m_{\uparrow}=m_{\downarrow}$, Here $k_{F}$ is the noninteracting
Fermi momentum for polarization $p = 0$.}\label{fig.1}
\end{figure}

\begin{figure}
\includegraphics[width=8.5cm, height=6.0cm]{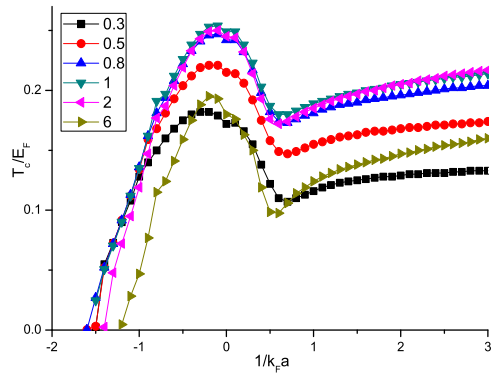}
\caption{(color online). Self-consistent solutions of the
transition temperature $T_{c}$ as a function of $1/k_{F}a$ for
different mass ratio $\eta  =m_{\uparrow}/m_{\downarrow}$ with the
polarization $p = 0$; $t = E_{F} = \hbar^{2} k_{F}^{2}/2 m$ and
$k_{F} dz = 1$. The transition temperature $T_{c}$ reduce as the
mass imbalance increasing.} \label{fig.2}
\end{figure}

\begin{figure}
\includegraphics[width=8.5cm, height=6.0cm]{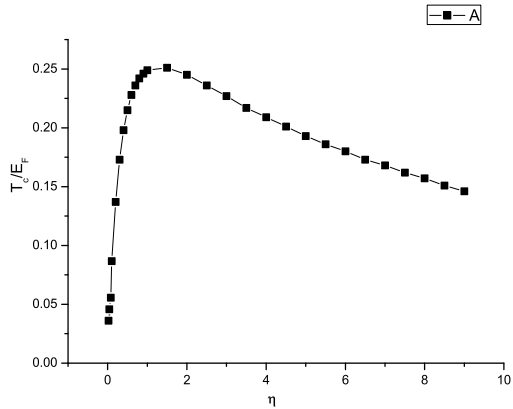}
\caption{ $T_{c}$ as a function of the mass ratio
$\eta=m_{\uparrow}/m_{\downarrow}$ at resonance $1/k_{F} a = 0$
with the polarization $p = 0$; $t = E_{F} = \hbar^{2} k_{F}^{2}/2
M$ and $k_{F} dz = 1$.} \label{fig.3}
\end{figure}

\begin{figure}
\includegraphics[width=8.5cm, height=6.0cm]{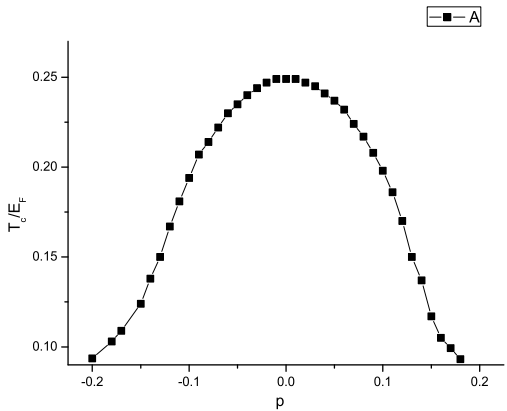}
\caption{ $T_{c}$ as a function of the polarization $p =
(n_{\uparrow} - n_{\downarrow})/n$ at resonance $1/k_{F} a = 0$ in
the case of $t = E_{F} = \hbar^{2} k_{F}^{2}/2 m$; $k_{F} dz = 1$
and $m_{\uparrow} = m_{\downarrow}$.} \label{fig.4}
\end{figure}

\section{Conclusion}

To summarize, we obtain the transition temperature of the
Fermi-Fermi mixture in Mix-dimensional system, as a function of
the interaction strength for different optical lattice parameter
$dz$ and mass ratio in the BCS-BEC crossover. In addition, we also
obtain the transition temperature behavior in resonance for
different mass ratio and polarization. We obtain the transition
temperature include the effect of the noncondensed cooper pairs by
the generalized mean field theory based on the BCS-Leggett ground
state in $G_{0}G$ schema. Increasing of the optical lattice
parameter $dz$ have positive effects to the transition
temperature. In BEC limit, the anisotropic effective mass
$M_{\perp} = 2 m$ for different $dz$. In mass case, the transition
temperature has maximum value at mass symmetry and reduce while
the masses deviate from the balance. Superfluid phase is
restricted within a narrow region of low polarization in
population imbalance case.

\section*{Acknowledgments}

We acknowledge useful discussions with Qijin Chen. This work is
supported by NSFC Grant No.10675108.

\section*{Appendix:}

In this appendix, we evaluate the expansion of the pair
susceptibility in low energy and long wavelength limit and derive
the explicit expression of the anisotropic effective mass of the
cooper pairs.

The pair susceptibility is given by
\begin{eqnarray}
\chi(Q) = \frac{1}{2}[\chi_{\uparrow \downarrow}(Q) +
\chi_{\downarrow \uparrow}(Q)] = \frac{1}{2} \sum_{K}[G_{0,
\uparrow}(Q - K) G_{\downarrow}(K) + G_{0, \downarrow}(Q - K)
G_{\uparrow}(K)],
\end{eqnarray}

The explicit expression of the pair susceptibility can be got by
substitute the Green's functions
\begin{eqnarray}
\chi(Q) &=& \frac{1}{2} \sum_{\textbf{k}}[\frac{1 -
f(\xi_{\textbf{k}}^{\beta}) - f(\xi_{\textbf{k} - \textbf{q},
\uparrow})} {\xi_{\textbf{k}}^{\beta} + \xi_{\textbf{k} -
\textbf{q}, \uparrow} - \imath \Omega_{n}} u_{\textbf{k}}^{2} -
\frac{f(\xi_{\textbf{k}}^{\alpha}) - f(\xi_{\textbf{k} -
\textbf{q}, \uparrow})}
{\xi_{\textbf{k}}^{\alpha} - \xi_{\textbf{k} - \textbf{q}, \uparrow} + \imath \Omega_{n}} v_{\textbf{k}}^{2} \nonumber\\
&&+ \frac{1 - f(\xi_{\textbf{k}}^{\alpha}) - f(\xi_{\textbf{k} -
\textbf{q}, \downarrow})} {\xi_{\textbf{k}}^{\alpha} +
\xi_{\textbf{k} - \textbf{q}, \downarrow} - \imath \Omega_{n}}
u_{\textbf{k}}^{2} - \frac{f(\xi_{\textbf{k}}^{\beta}) -
f(\xi_{\textbf{k} - \textbf{q}, \downarrow})}
{\xi_{\textbf{k}}^{\alpha} - \xi_{\textbf{k} - \textbf{q},
\downarrow} + \imath \Omega_{n}} v_{\textbf{k}}^{2}],
\end{eqnarray}

At transition temperature $T_{c}$, $t_{sc}^{-1}(Q) = 0$ and
\begin{eqnarray}
t_{pg}^{-1}(Q) = t^{-1}(Q) = \chi(Q) - \chi(0),
\end{eqnarray}

In low energy and long wavelength limit
\begin{eqnarray}
t_{pg}^{-1}(Q) = Z(\imath\Omega_{n} - \Omega_{\textbf{q}} +
\mu_{pair} + \imath \Gamma_{q, \Omega}),
\end{eqnarray}

$\Omega_{\textbf{q}}=\frac{\textbf{q}_{\bot}^{2}}{2M_{\perp}^{*}}+\frac{\textbf{q}_{z}^{2}}{2M_{z}^{*}}$
is the dispersion of the cooper pairs. $M_{z}$ and $M_{\perp}$ is
the anisotropic effective mass. Below the critical temperature
$T_{c}$ the chemical potential of cooper pair vanish
$\mu_{pair}=0$ as for the BEC condition. The imaginary part of the
pair dispersion is given by

\begin{eqnarray}
\Gamma_{q, \Omega} &=& \frac{\pi}{Z} \sum_{\textbf{k}}\{[1 -
f(\xi_{\textbf{k}}^{\beta}) - f(\xi_{\textbf{k} - \textbf{q},
\uparrow})] u_{\textbf{k}}^{2} \delta(\xi_{\textbf{k}}^{\beta} +
\xi_{\textbf{k} - \textbf{q}, \uparrow} - \Omega) +
[f(\xi_{\textbf{k}}^{\alpha}) - f(\xi_{\textbf{k} - \textbf{q},
\uparrow})] v_{\textbf{k}}^{2} \delta(\xi_{\textbf{k}}^{\alpha} - \xi_{\textbf{k} - \textbf{q}, \uparrow} + \Omega) \nonumber\\
&&+ [1 - f(\xi_{\textbf{k}}^{\alpha}) - f(\xi_{\textbf{k} -
\textbf{q}, \downarrow})] u_{\textbf{k}}^{2}
\delta(\xi_{\textbf{k} - \textbf{q}, \downarrow} - \Omega) +
[f(\xi_{\textbf{k}}^{\beta}) - f(\xi_{\textbf{k} - \textbf{q},
\downarrow})] v_{\textbf{k}}^{2} \delta(\xi_{\textbf{k}}^{\alpha}
- \xi_{\textbf{k} - \textbf{q}, \downarrow} + \Omega)\},
\end{eqnarray}
and the reverse of the residue is
\begin{eqnarray}
Z &=& \frac{\partial \chi}{\partial \imath \Omega} \mid_{\Omega = \textbf{q} = 0} \nonumber\\
&=& \frac{1}{2 \Delta^{2}}[n -
\Sigma_{\textbf{k}}(f(\xi_{\textbf{k}, \uparrow}) +
f(\xi_{\textbf{k}, \downarrow}))],
\end{eqnarray}

The anisotropic effective masses of the cooper pair are given by
\begin{eqnarray*}
\frac{1}{2M_{\perp}^{*}} &=& - \frac{1}{4 Z} \frac{\partial^{2} t_{pg}^{-1}}{\partial \textbf{q}_{\perp}^{2}} \mid_{\Omega = \textbf{q} = 0} \\
&=& - \frac{1}{8 Z} \sum_{\textbf{k}} \{\frac{2
f'(\xi_{\textbf{k}, \uparrow})}{\Delta^{2}} (\frac{\partial
\xi_{\textbf{k}, \uparrow}}{\partial \textbf{k}_{\perp}})^{2} +
\frac{2f'(\xi_{\textbf{k}, \downarrow})}
{\Delta^{2}}(\frac{\partial \xi_{\textbf{k}, \downarrow}}{\partial \textbf{k}_{\perp}})^{2} \\
&&- \frac{(1 - f(\xi_{\textbf{k}}^{\beta}) -
f(\xi_{\textbf{k}}^{\uparrow}))(E_{\textbf{k}} -
\xi_{\textbf{k}}^{+}) + (f(\xi_{\textbf{k}}^{\alpha}) -
f(\xi_{\textbf{k}}^{\uparrow}))(E_{\textbf{k}} +
\xi_{\textbf{k}}^{+})}
{2 E_{\textbf{k}} \Delta^{2}}(\frac{\partial^{2} \xi_{\textbf{k}, \uparrow}}{\partial \textbf{k}_{\bot}^{2}}) \\
&&- \frac{(1 - f(\xi_{\textbf{k}}^{\alpha}) -
f(\xi_{\textbf{k}}^{\downarrow}))(E_{\textbf{k}} -
\xi_{\textbf{k}}^{+}) + (f(\xi_{\textbf{k}}^{\beta}) -
f(\xi_{\textbf{k}}^{\downarrow}))(E_{\textbf{k}} +
\xi_{\textbf{k}}^{+})}
{2 E_{\textbf{k}} \Delta^{2}}(\frac{\partial^{2} \xi_{\textbf{k}, \downarrow}}{\partial \textbf{k}_{\bot}^{2}}) \\
&&+ \frac{(1 - f(\xi_{\textbf{k}}^{\beta}) -
f(\xi_{\textbf{k}}^{\uparrow}))(E_{\textbf{k}} -
\xi_{\textbf{k}}^{+})^{2} - (f(\xi_{\textbf{k}}^{\alpha}) -
f(\xi_{\textbf{k}}^{\uparrow}))(E_{\textbf{k}} +
\xi_{\textbf{k}}^{+})^{2}}{E_{\textbf{k}} \Delta^{4}}(\frac{\partial \xi_{\textbf{k}, \uparrow}}{\partial \textbf{k}_{\bot}})^{2} \\
&&+ \frac{(1 - f(\xi_{\textbf{k}}^{\alpha}) -
f(\xi_{\textbf{k}}^{\downarrow}))(E_{\textbf{k}} -
\xi_{\textbf{k}}^{+})^{2} - (f(\xi_{\textbf{k}}^{\beta}) -
f(\xi_{\textbf{k}}^{\downarrow}))(E_{\textbf{k}} +
\xi_{\textbf{k}}^{+})^{2}} {E_{\textbf{k}}
\Delta^{4}}(\frac{\partial \xi_{\textbf{k}, \downarrow}}{\partial
\textbf{k}_{\bot}})^{2}\},
\end{eqnarray*}
\begin{eqnarray*}
\frac{1}{2M_{z}^{*}} &=& - \frac{1}{2 Z} \frac{\partial^{2} t_{pg}^{-1}}{\partial k_{z}^{2}} \mid_{\Omega = \textbf{q} = 0} \\
&=&- \frac{1}{4 Z} \sum_{\textbf{k}}\{\frac{2 f'(\xi_{\textbf{k},
\uparrow})}{\Delta^{2}} (\frac{\partial \xi_{\textbf{k},
\uparrow}}{\partial k_{z}})^{2} + \frac{2 f'(\xi_{\textbf{k},
\downarrow})} {\Delta^{2}}(\frac{\partial \xi_{\textbf{k},
\downarrow}}{\partial k_{z}})^{2} \\
&&- \frac{(1 - f(\xi_{\textbf{k}}^{\beta}) -
f(\xi_{\textbf{k}}^{\uparrow}))(E_{\textbf{k}} -
\xi_{\textbf{k}}^{+}) + (f(\xi_{\textbf{k}}^{\alpha}) -
f(\xi_{\textbf{k}}^{\uparrow}))(E_{\textbf{k}} +
\xi_{\textbf{k}}^{+})} {2 E_{\textbf{k}} \Delta^{2}}
\frac{\partial^{2} \xi_{\textbf{k}, \uparrow}}{\partial k_{z}^{2}} \\
&&- \frac{(1 - f(\xi_{\textbf{k}}^{\alpha}) -
f(\xi_{\textbf{k}}^{\downarrow}))(E_{\textbf{k}} -
\xi_{\textbf{k}}^{+}) + (f(\xi_{\textbf{k}}^{\beta}) -
f(\xi_{\textbf{k}}^{\downarrow}))(E_{\textbf{k}} +
\xi_{\textbf{k}}^{+})} {2 E_{\textbf{k}} \Delta^{2}}
\frac{\partial^{2} \xi_{\textbf{k}, \downarrow}}{\partial k_{z}^{2}} \\
&&+ \frac{(1 - f(\xi_{\textbf{k}}^{\beta}) -
f(\xi_{\textbf{k}}^{\uparrow}))(E_{\textbf{k}} -
\xi_{\textbf{k}}^{+})^{2} - (f(\xi_{\textbf{k}}^{\alpha}) -
f(\xi_{\textbf{k}}^{\uparrow}))(E_{\textbf{k}} +
\xi_{\textbf{k}}^{+})^{2}} {E_{\textbf{k}}
\Delta^{4}}(\frac{\partial \xi_{\textbf{k}, \uparrow}}{\partial k_{z}})^{2} \\
&&+ \frac{(1 - f(\xi_{\textbf{k}}^{\alpha}) -
f(\xi_{\textbf{k}}^{\downarrow}))(E_{\textbf{k}} -
\xi_{\textbf{k}}^{+})^{2} - (f(\xi_{\textbf{k}}^{\beta}) -
f(\xi_{\textbf{k}}^{\downarrow}))(E_{\textbf{k}} +
\xi_{\textbf{k}}^{+})^{2}} {E_{\textbf{k}}
\Delta^{4}}(\frac{\partial \xi_{\textbf{k}, \downarrow}}{\partial
k_{z}})^{2}\},
\end{eqnarray*}

\end{document}